\documentclass[oribibl,envcountsame]{llncs}
\usepackage[ps,matrix,arrow,curve]{xypic}
\usepackage[psamsfonts]{amsfonts}
\usepackage{amsfonts,amssymb,stmaryrd,amsmath}
\usepackage{epsfig}
\usepackage{url}
\usepackage{galois}
\usepackage{gastex}
\usepackage{color}

\xyoption{xdvi}

\newcommand{\mydef}[1]{{\em #1}}

\newenvironment{mytheoreml}
      {\begin{theorem}~\par\begin{enumerate}}
      {\end{enumerate}\hfill$\Box$\end{theorem}}

\newenvironment{mydefinitionl}
      {\begin{definition}~\par\begin{enumerate}}
      {\end{enumerate}\hfill$\Box$\end{definition}}

\newenvironment{mytheorem}
      {\begin{theorem}~\par}
      {\hfill$\Box$\end{theorem}}

\newcommand{\deltaeq}{\;\stackrel{\triangle}{=}\;}
\newcommand{\deltaiff}{\stackrel{\triangle}{\iff}}

\def\doframeit#1{\vbox{%
  \hrule height\fboxrule
    \hbox{%
      \vrule width\fboxrule \kern\fboxsep
      \vbox{\kern\fboxsep #1\kern\fboxsep }%
      \kern\fboxsep \vrule width\fboxrule }%
    \hrule height\fboxrule }}

\def\frameit{\smallskip \advance \linewidth by -7.5pt \setbox0=\vbox \bgroup
\strut \ignorespaces }

\def\endframeit{\ifhmode \par \nointerlineskip \fi \egroup
\doframeit{\box0}}

\begin{document}


\title{A New Numerical Abstract Domain \\ Based on Difference-Bound Matrices}
\titlerunning{A New Numerical Abstract Domain Based on DBMs}

\author{Antoine Min\'e}
\institute{\'Ecole Normale Sup\'erieure de Paris, France,\\
           \email{mine@di.ens.fr},\\
           \url{http://www.di.ens.fr/~mine}}
\tocauthor{Antoine Min\'e (ENS Paris)}

\maketitle



\begin{abstract}
This paper presents a new numerical abstract domain for 
static analysis by abstract interpretation. This domain allows us to represent
invariants of the form $(x-y\leq c)$
and $(\pm x\leq c)$, where $x$ and $y$ are variables values and $c$ is an
integer or real constant.

Abstract elements are represented by Difference-Bound Matrices,
widely used by model-checkers, but we had to design new operators to meet
the needs of abstract interpretation.
The result is a complete lattice of infinite height
featuring widening, narrowing and common transfer functions.

We focus on giving an efficient $\mathcal{O}(n^2)$ representation
and graph-based $\mathcal{O}(n^3)$ algorithms---where $n$ is the number of 
variables---and claim that
this domain always performs more precisely than the well-known interval 
domain.

To illustrate the precision/cost tradeoff of this domain, we have
implemented simple abstract interpreters for toy imperative and
parallel languages which allowed us to prove some non-trivial algorithms
correct.
\end{abstract}


\section{Introduction}

Abstract interpretation has proved to be a useful tool for eliminating
bugs in software because it allows the design
of automatic and sound analyzers for real-life programming languages.
While abstract interpretation is a very general framework, we will be
interested here only in discovering numerical invariants, that is to say,
arithmetic relations that hold between numerical variables in a program.
Such invariants are useful for tracking common errors such as
division by zero and out-of-bound array access.

In this paper we propose practical algorithms to discover invariants
of the form $(x-y\leq c)$ and $(\pm x\leq c)$---where $x$ and $y$ are
numerical program variables and $c$ is a numeric constant.
Our method works for integers, reals and even rationals.

For the sake of brevity, we will omit proofs of theorems in this paper.
The complete proof for all theorems can be found in 
the author's MS thesis \cite{dea}.

\subsubsection{Previous and Related Work.}
Static analysis has developed approaches to automatically find numerical 
invariants based on \mydef{numerical abstract domains} 
representing the form of the invariants we want to find.
Famous examples are the lattice of intervals 
(described in, for instance, Cousot and Cousot's ISOP'76 paper
\cite{interv}) and the lattice of polyhedra 
(described in Cousot and Halbwachs's POPL'78 paper \cite{poly}) 
which represent respectively 
invariants of the form
$(v\in[c_1,c_2])$ and $(\alpha_1v_1+\cdots+\alpha_nv_n\leq c)$.
Whereas the interval analysis is very efficient---linear 
memory and time cost---but not very precise, the
polyhedron analysis is much more precise but has a huge memory 
cost---exponential in the number of variables.

Invariants of the form $(x-y\leq c)$ and $(\pm x\leq c)$
are widely used by the {\em model-checking} community.
A special representation, called \mydef{Difference-Bound Matrices (DBMs)}, 
was introduced, as well as many operators 
in order to model-check {\em timed automata}
(see Yovine's ES'98 paper \cite{DBM2} and
Larsen, Larsson, Pettersson and Yi's RTSS'97 paper \cite{DBM}).
Unfortunately, most operators are tied to model-checking and are 
of little interest for static analysis.

\subsubsection{Our Contribution.}
This paper presents a new abstract numerical domain
based on the DBM representation,
together with a full set of new operators and transfer functions adapted
to static analysis.

Sections 2 and 3 present a few well-known results about potential constraint
sets and introduce briefly the Difference-Bound Matrices.
Section 4 presents operators
and transfer functions that are new---except for the intersection 
operator---and adapted to abstract interpretation.
In Section 5, we use these operators to build {\em lattices}, which can
be complete under certain conditions.
Section 6 shows some practical results we obtained with an example
implementation and Section 7 gives some ideas for improvement.

\section{Difference-Bound Matrices}

Let $\mathcal{V}=\{v_1,\ldots,v_n\}$ be a finite set a variables with value
in a numerical set $\mathbb{I}$ (which can be the set $\mathbb{Z}$ 
of integers, the set $\mathbb{Q}$ of rationals or
the set $\mathbb{R}$ of reals).

We focus, in this paper, on the representation of constraints of the form
$(v_j-v_i\leq c)$, $(v_i\leq c)$ and $(v_i\geq c)$, 
where $v_i,v_j\in\mathcal{V}$ and $c\in\mathbb{I}$.
By choosing one variable to be always equal to $0$, we can represent the
above constraints using only \mydef{potential constraints}, that is
to say, constraints of the form $(v_j-v_i\leq c)$.
From now, we will choose $v_2, \ldots, v_n$ to be program
variables, and $v_1$ to be the constant $0$ so that
$(v_i\leq c)$ and $(v_i\geq c)$ are 
rewritten $(v_i-v_1\leq c)$ and $(v_1-v_i\leq -c)$.
We assume we now work only with potential constraints over 
the set $\{v_1,\ldots,v_n\}$.

\subsubsection{Difference-Bound Matrices.} 
We extend $\mathbb{I}$ to $\overline{\mathbb{I}}=\mathbb{I}\cup\{+\infty\}$ 
by adding the $+\infty$ element. 
The standard operations $\leq$, $=$, $+$, $\min$ and $\max$ are
extended to $\overline{\mathbb{I}}$ as usual (we will not use
operations, such as $-$ or $*$, that may lead to indeterminate forms).

Any set $C$ of potential constraints over $\mathcal{V}$
can be represented uniquely by a $n\times n$ matrix in 
$\overline{\mathbb{I}}$---provided we assume, without loss of generality, that 
there does not exist two potential constraints $(v_j-v_i\leq c)$ in $C$
with the same left member and different right members.
The matrix $\vec{m}$ associated with the potential constraint set $C$ is
called a \mydef{Difference-Bound Matrix (DBM)} and is defined as follows:
$$
\vec{m}_{ij} \deltaeq \left\{
\begin{array}{ll}
c&\mbox{ if }(v_j-v_i\leq c)\in C,\\
+\infty\quad&\mbox{ elsewhere}\enspace.
\end{array}
\right.
$$

\subsubsection{Potential Graphs.}
A DBM $\vec{m}$ can be seen as the adjacency matrix of a directed graph 
$\mathcal{G}=(\mathcal{V},\mathcal{A},w)$
with edges weighted in $\mathbb{I}$.
$\mathcal{V}$ is the set of nodes, $\mathcal{A}\subseteq\mathcal{V}^2$ 
is the set of edges and $w\in\mathcal{A}\mapsto\mathbb{I}$ is the weight 
function. $\mathcal{G}$ is defined by:
$$\left\{\begin{array}{ll}
(v_i,v_j)\notin\mathcal{A}
 &\mbox{ if }\vec{m}_{ij}=+\infty,\\
(v_i,v_j)\in\mathcal{A}\mbox{ and }w(v_i,v_j)=\vec{m}_{ij}\quad
 &\mbox{ if }\vec{m}_{ij}\neq +\infty\enspace.
\end{array}\right.$$

We will denote by 
$\langle i_1,\ldots,i_k \rangle$ a {\em finite} set of
nodes representing a \mydef{path} from node $v_{i_1}$ to node $v_{i_k}$ 
in $\mathcal{G}$.
A \mydef{cycle} is a path such that $i_1=i_k$.

\subsubsection{$\mathcal{V}$-Domain and $\mathcal{V}^0$-Domain.}
We call the \mydef{$\mathcal{V}$-domain} of a DBM $\vec{m}$ 
and we denote by $\mathcal{D}(\vec{m})$ the set
of points in $\mathbb{I}^n$ that satisfy all potential constraints:
$$
\mathcal{D}(\vec{m})\deltaeq\{(x_1,\ldots,x_n)\in\mathbb{I}^n
\;|\;\forall i,j,\; x_j-x_i\leq \vec{m}_{ij}\}\enspace.
$$

Now, remember that the variable $v_1$ has a special semantics: it is always 
equal to $0$.
Thus, it is not the $\mathcal{V}$-domain which is of interest, but
the \mydef{$\mathcal{V}^0$-domain}
(which is a sort of
intersection-projection of the $\mathcal{V}$-domain)
denoted by $\mathcal{D}^0(\vec{m})$ and defined by:
$$
\mathcal{D}^0(\vec{m})\deltaeq 
\{(x_2,\ldots,x_n)\in\mathbb{I}^{n-1}\;|\;
(0,x_2,\ldots,x_n)\in \mathcal{D}(\vec{m})\}\enspace.
$$

We will call \mydef{$\mathcal{V}$-domain} and
\mydef{$\mathcal{V}^0$-domain} 
any subset of $\mathbb{I}^n$ or $\mathbb{I}^{n-1}$
which is respectively the $\mathcal{V}$-domain or the $\mathcal{V}^0$-domain
of some DBM. Figure \ref{domainfig} shows an example DBM together with its
corresponding potential graph, constraint set, $\mathcal{V}$-domain and 
$\mathcal{V}^0$-domain.

\begin{figure}
\begin{frameit}
\begin{center}
\begin{tabular}{c}
\begin{tabular}{cccccc}
\\
(a)&
$
\left\{
\begin{array}{rcr}
v_2&\leq& 4\\
-v_2&\leq& -1\\
v_3&\leq& 3\\
-v_3&\leq& -1\\
v_2-v_3&\leq& 1
\end{array}
\right.
$

&\quad(b)&
$
\begin{array}{c|ccc}
& v_1 & v_2 & v_3\\
\hline
v_1 & +\infty & 4 & 3 \\
v_2 & -1 & +\infty & +\infty\\
v_3 & -1 & 1 & +\infty\\
\end{array}
$

&\quad(c)&
\raisebox{0.5cm}{
\xymatrix{&
*++[o][F-]{v_1} \ar@/_1pc/[ld]_4 \ar[rd]_3 & \\
*++[o][F-]{v_2} \ar[ur]_{-1} & & 
*++[o][F-]{v_3} \ar@/_1pc/[ul]_{-1} \ar[ll]^1
}}

\end{tabular}
\\

\begin{tabular}{cccc}

(d)&
\raisebox{-1.5cm}{\input{vdomain.pstex_t}}\quad

&\quad(e)&

\raisebox{-1.5cm}{\input{v0domain.pstex_t}}
\\
\end{tabular}
\end{tabular}
\end{center}
\caption{A constraint set (a), its corresponding DBM (b) and
potential graph (c), its $\mathcal{V}$-domain (d) and
$\mathcal{V}^0$-domain (e).}
\label{domainfig}
\end{frameit}
\end{figure}

\subsubsection{$\trianglelefteqslant$ Order.}
The $\leq$ order on $\overline{\mathbb{I}}$ induces a point-wise
order $\trianglelefteqslant$ on the set of DBMs:
$$
\vec{m}\trianglelefteqslant\vec{n}\deltaiff\forall i,j,\;\vec{m}_{ij}\leq \vec{n}_{ij}\enspace.
$$
This order is {\em partial}. It is also {\em complete} if $\mathbb{I}$
has least-upper bounds, i.e, if $\mathbb{I}$ is $\mathbb{R}$ or $\mathbb{Z}$, 
but not $\mathbb{Q}$.
We will denote by $=$ the associated equality relation which is simply
the matrix equality.

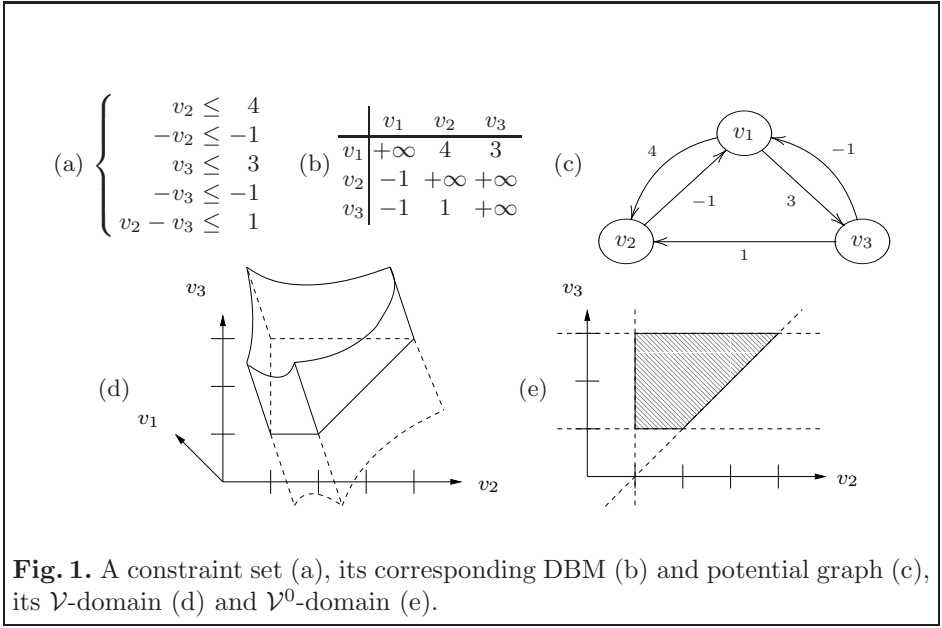
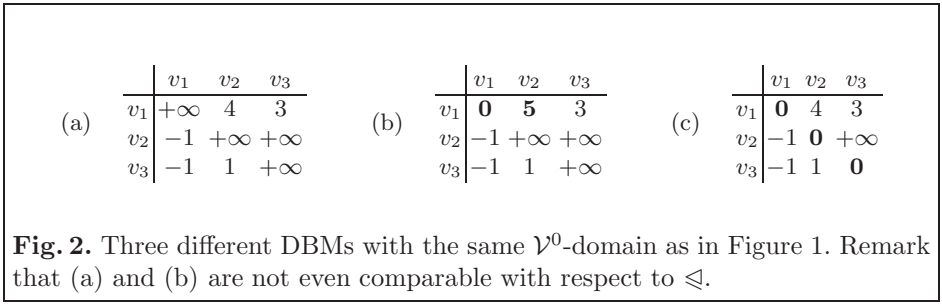
\begin{figure}
\begin{frameit}
\begin{center}
\begin{tabular}{cccccc}
(a)&\quad
$
\begin{array}{c|ccc}
& v_1 & v_2 & v_3\\
\hline
v_1 & +\infty & 4 & 3 \\
v_2 & -1 & +\infty & +\infty\\
v_3 & -1 & 1 & +\infty\\
\end{array}
$
\quad\quad\quad
&
(b)&\quad
$
\begin{array}{c|ccc}
& v_1 & v_2 & v_3\\
\hline
v_1 & {\bf 0} & {\bf 5} & 3 \\
v_2 & -1 & +\infty & +\infty\\
v_3 & -1 & 1 & +\infty\\
\end{array}
$
\quad\quad\quad
&
(c)&\quad
$
\begin{array}{c|ccc}
& v_1 & v_2 & v_3\\
\hline
v_1 & {\bf 0} & 4 & 3 \\
v_2 & -1 & {\bf 0} & +\infty\\
v_3 & -1 & 1 & {\bf 0}\\
\end{array}
$
\end{tabular}
\end{center}
\caption{Three different DBMs with the same $\mathcal{V}^0$-domain as in 
Figure \ref{domainfig}.
Remark that (a) and (b) are not even
comparable with respect to $\trianglelefteqslant$.
}
\label{orderfig}
\end{frameit}
\end{figure}

We have $\vec{m}\trianglelefteqslant\vec{n}\;\Longrightarrow\;
\mathcal{D}^0(\vec{m})\subseteq\mathcal{D}^0(\vec{n})$
but the converse
is not true.
In particular, we {\em do not have} $\mathcal{D}^0(\vec{m})=
\mathcal{D}^0(\vec{n})\;\Longrightarrow\;\vec{m}=\vec{n}$
(see Figure \ref{orderfig} for a counter-example).

\section{Closure, Emptiness, Inclusion and Equality Tests}

We saw in Figure \ref{orderfig} that two different DBMs can represent
the same $\mathcal{V}^0$-domain.
In this section, we show that there exists a normal form for any
DBM with a non-empty $\mathcal{V}^0$-domain
and present an algorithm to find it.
The existence and computability of a normal form is very important
since it is, as often in abstract representations, 
the key to equality testing used in fixpoint computation.
In the case of DBMs, it will also allows us to carry an analysis of the
precision of the operators defined in the next section.

\subsubsection{Emptiness Testing.}
We have the following graph-oriented theorem:

\begin{mytheorem}
\label{emptythm}
\item A DBM has an empty $\mathcal{V}^0$-domain if and only if there exists,
in its associated potential graph, 
a cycle with a strictly negative total weight.
\end{mytheorem}
Checking for cycles with a strictly negative weight is done using the
well-known {\em Bellman-Ford algorithm} which runs in
$\mathcal{O}(n^3)$. This algorithm
can be found in Cormen, Leiserson and Rivest's classical 
algorithmics textbook \cite[\S 25.3]{CLR}.


\subsubsection{Closure and Normal Form.}
Let $\vec{m}$ be a DBM {\em with a non-empty $\mathcal{V}^0$-domain}
and $\mathcal{G}$ its associated potential graph. Since $\mathcal{G}$ has
no cycle with a strictly negative weight, we can compute its 
{\em shortest path closure} $\mathcal{G}^*$, 
the adjacency matrix of which will be denoted by $\vec{m}^*$ and defined by:
$$
\left\{\begin{array}{ll}
\vec{m}^*_{ii}\deltaeq 0,&\\
\displaystyle
\vec{m}^*_{ij}\deltaeq\min_{\substack{
1\leq N\\ 
\langle i=i_1,i_2,\ldots, i_N=j\rangle\quad}}
\sum_{k=1}^{N-1} \vec{m}_{i_k i_{k+1}}\quad&\mbox{if }i\neq j\enspace.
\end{array}
\right.
$$
The idea of closure relies on the fact that,
if $\langle i=i_1,i_2,\ldots,i_N=j \rangle$
is a path from $v_i$ to $v_j$, then the constraint
$v_j-v_i \leq \sum_{k=1}^{N-1} \vec{m}_{i_k i_{k+1}}$ can be derived from $\vec{m}$
by adding the potential constraints 
$v_{i_{k+1}}-v_{i_k} \leq \vec{m}_{i_k i_{k+1}}, 1\leq k\leq N-1$.
This is an \mydef{implicit} potential constraint which does not appear 
directly in the DBM $\vec{m}$.
When computing the closure, we replace each potential constraint 
$v_j-v_i\leq \vec{m}_{ij}, i\neq j$
in $\vec{m}$  by the tightest implicit constraint we can find, and
each diagonal element by $0$ (which is indeed the smallest value
$v_i-v_i$ can reach).
In Figure \ref{orderfig} for instance, (c) is the closure of 
both the (a) and (b) DBMs.

\begin{mytheoreml}
\label{closurethm}
\item $\vec{m}^*=\inf_{\trianglelefteqslant}\{\vec{n}\;|\;
\mathcal{D}^0(\vec{n})=\mathcal{D}^0(\vec{m})\}$.
\item $\mathcal{D}^0(\vec{m})$ {\em saturates} $\vec{m}^*$, that is to say:\\
$\forall i,j,\mbox{ such that }\vec{m}^*_{ij}<+\infty,\;\exists 
(x_1=0,x_2,\ldots, x_n)\in\mathcal{D}(\vec{m}),\;x_j-x_i=\vec{m}^*_{ij}$.
\end{mytheoreml}

Theorem \ref{closurethm}.1 states that $\vec{m}^*$ is the smallest 
DBM---with respect to $\trianglelefteqslant$---that represents a given
$\mathcal{V}^0$-domain, and thus {\it the closed form is a normal form}.
Theorem \ref{closurethm}.2 is a crucial property to prove accuracy of some
operators defined in the next section.

Any shortest-path graph algorithm can be used to compute the closure
of a DBM.
We suggest the straightforward {\em Floyd-Warshall},
which is described in Cormen, Leiserson and Rivest's textbook
\cite[\S 26.2]{CLR}, and has a $\mathcal{O}(n^3)$ time cost.


\subsubsection{Equality and Inclusion Testing.}
The case where $\vec{m}$ or $\vec{n}$ or both have an 
empty $\mathcal{V}^0$-domain is easy; in all other cases we use
the following theorem---which is a consequence of Theorem \ref{closurethm}.1:
\begin{mytheoreml}
\label{testthm}
\item If $\vec{m}$ and $\vec{n}$ have non-empty $\mathcal{V}^0$-domain,
$\mathcal{D}^0(\vec{m})=\mathcal{D}^0(\vec{n}) \iff
\vec{m}^*=\vec{n}^*$.
\item If $\vec{m}$ and $\vec{n}$ have non-empty $\mathcal{V}^0$-domain,
$\mathcal{D}^0(\vec{m})\subseteq\mathcal{D}^0(\vec{n}) \iff
\vec{m}^*\trianglelefteqslant \vec{n}$.
\end{mytheoreml}
Besides emptiness test and closure, we may need, in order to
test equality or inclusion, to compare matrices with respect to the
point-wise ordering $\trianglelefteqslant$. This can be done with a
$\mathcal{O}(n^2)$ time cost.

\subsubsection{Projection.}
We define the \mydef{projection} $\pi_{|v_k}(\vec{m})$ of a DBM $\vec{m}$
with respect to a variable $v_k$ to be the interval containing
all possible values of $v\in\mathbb{I}$ such that there exists a point
$(x_2,\ldots,x_n)$ in the $\mathcal{V}^0$-domain of $\vec{m}$
with $x_k=v$:
$$
\pi_{|v_k}(\vec{m})\deltaeq\{x\in\mathbb{I}\;|\;\exists(x_2,\ldots,x_n)\in
\mathcal{D}^0(\vec{m})\mbox{ such that }x=x_k\}\enspace.
$$
The following theorem, which is a consequence of the saturation property of 
the closure, gives an algorithmic way to compute the projection:
\begin{mytheorem}
\label{projectionthm}
If $\vec{m}$ has a non-empty $\mathcal{V}^0$-domain, then
$\pi_{|v_k}(\vec{m})=\left[-\vec{m}^*_{k1},\vec{m}^*_{1k}\right]$\\
(interval bounds are included only if finite).
\end{mytheorem}

\section{Operators and Transfer Functions}
In this section, we define some operators and transfer functions to be
used in abstract semantics.
Except for the intersection operator, they are new.
The operators are basically
point-wise extensions of the standard operators
defined over the domain of intervals \cite{interv}.

Most algorithms presented here are either constant time, or
point-wise, i.e., quadratic time.


\subsubsection{Intersection.}
Let us define the point-wise 
\mydef{intersection DBM $\vec{m}\wedge \vec{n}$} by:
$$
(\vec{m}\wedge \vec{n})_{ij}\deltaeq\min(\vec{m}_{ij},\vec{n}_{ij})\enspace .
$$
We have the following theorem:
\begin{mytheorem}
\label{intersectionthm}
$\mathcal{D}^0(\vec{m}\wedge \vec{n})=
\mathcal{D}^0(\vec{m})\cap\mathcal{D}^0(\vec{n})$.
\end{mytheorem}
stating that the intersection is always exact.
However, the resulting DBM is seldom closed, even if the arguments are closed.

\subsubsection{Least Upper Bound.}
The set of $\mathcal{V}^0$-domains is not stable by union\footnote{
$\mathcal{V}^0$-domains are always convex, but the union of two
$\mathcal{V}^0$-domains may not be convex.} so we introduce here 
a union operator which over-approximate its result.
We define the point-wise \mydef{least upper bound DBM $\vec{m}\vee\vec{n}$}
by:
$$
(\vec{m}\vee \vec{n})_{ij}\deltaeq\max(\vec{m}_{ij},\vec{n}_{ij})\enspace .
$$

$\vec{m}\vee\vec{n}$ is indeed the least upper bound with respect
to the $\trianglelefteqslant$ order.
The following theorem tells us about the effect of this operator on
$\mathcal{V}^0$-domains:
\begin{mytheoreml}
\label{lubthm}
\item $\mathcal{D}^0(\vec{m}\vee \vec{n})\supseteq
\mathcal{D}^0(\vec{m})\cup\mathcal{D}^0(\vec{n})$.
\item If $\vec{m}$ and $\vec{n}$ have non-empty $\mathcal{V}^0$-domains, then
$$(\vec{m}^*)\vee(\vec{n}^*)=\inf_{\trianglelefteqslant}
\{\vec{o}\;|\; \mathcal{D}^0(\vec{o})\supseteq\mathcal{D}^0(\vec{m})
\cup\mathcal{D}^0(\vec{n})\}$$ and, as a consequence, 
$\mathcal{D}^0((\vec{m}^*)\vee(\vec{n}^*))$
is the smallest $\mathcal{V}^0$-domain (with respect to the $\subseteq$
ordering) which contains $\mathcal{D}^0(\vec{m})\cup\mathcal{D}^0(\vec{n})$.
\item If $\vec{m}$ and $\vec{n}$ are closed, then so is
$\vec{m}\vee\vec{n}$.
\end{mytheoreml}
Theorem \ref{lubthm}.1 states that $\mathcal{D}^0(\vec{m}\vee \vec{n})$ is
an upper bound in the set of $\mathcal{V}^0$-domains with respect to the
$\subseteq$ order.
If precision is a concern, we need to find the {\em least} upper bound
in this set. Theorem \ref{lubthm}.2---which is a consequence of the
saturation property of the closure---states 
that {\em we have to close both arguments} before applying the $\vee$ 
operator to get this most precise union over-approximation. 
If one argument has an empty 
$\mathcal{V}^0$-domain, the least upper bound we want is simply the
other argument. Emptiness tests and closure add a $\mathcal{O}(n^3)$
time cost.


\subsubsection{Widening.}
When computing the semantics of a program, one often encounters {\em loops}
leading to fixpoint computation involving infinite iteration sequences.
In order to compute {\em in finite time} an upper approximation of a fixpoint, 
\mydef{widening operators}
were introduced in P. Cousot's thesis \cite[\S 4.1.2.0.4]{these}.
Widening is a sort of union for which every increasing chain is
stationary after a finite number of iterations.
We define the point-wise widening operator $\triangledown$ by:
$$(\vec{m}\triangledown \vec{n})_{ij}\deltaeq
\left\{
\begin{array}{ll}
\vec{m}_{ij} & \mbox{if }\vec{n}_{ij}\leq \vec{m}_{ij},\\
+\infty \quad& \mbox{elsewhere}\enspace.
\end{array}
\right.
$$
The following properties prove that $\triangledown$ is
indeed a widening:
\begin{mytheoreml}
\label{wideningthm}
\item 
$\mathcal{D}^0(\vec{m}\triangledown\vec{n})\supseteq\mathcal{D}^0(\vec{m})\cup\mathcal{D}^0(\vec{n})$.
\item {\em Finite chain property}:

$\forall \vec{m}$ and $\forall (\vec{n_i})_{i\in\mathbb{N}}$, 
the chain defined by:
$$
\left\{
\begin{array}{lcl}
\vec{x_0}&\deltaeq&\vec{m},\\
\vec{x_{i+1}}&\deltaeq&\vec{x_i}\triangledown\vec{n_i},
\end{array}
\right.
$$
is increasing for $\trianglelefteqslant$ and ultimately stationary.
The limit $\vec{l}$ is such that $\vec{l}\trianglerighteqslant\vec{m}$ and
$\forall i,\;\vec{l}\trianglerighteqslant\vec{n_i}$.
\end{mytheoreml}
The widening operator has some intriguing interactions with closure.
Like the least upper bound, the widening operator gives more precise results
if its right argument is closed, so it is
rewarding to change $\vec{x_{i+1}}=\vec{x_i}\triangledown\vec{n_i}$ into
$\vec{x_{i+1}}=\vec{x_i}\triangledown(\vec{n_i}^*)$.
This is not the case for the first argument: we can have sometimes
$\mathcal{D}^0(\vec{m}\triangledown\vec{n})\varsubsetneq
\mathcal{D}^0((\vec{m^*})\triangledown\vec{n})$.
Worse, if we try to force the closure of the first argument by changing
$\vec{x_{i+1}}=\vec{x_i}\triangledown\vec{n_i}$ into
$\vec{x_{i+1}}=(\vec{x_i}\triangledown\vec{n_i})^*$, the finite chain
property (Theorem \ref{wideningthm}.2) {\em is no longer satisfied},
as illustrated in Figure \ref{wideningfig}.

\begin{figure}
\begin{frameit}
\begin{center}
\begin{tabular}{cc}
$\vec{m}\deltaeq$
\raisebox{0.75cm}{
\xymatrix{& 
*++[o][F-]{v_1} \ar[dl]^{       1} & \\
*++[o][F-]{v_2} \ar@/^1pc/[ur]^{1} \ar[rr]^{       1} & & 
*++[o][F-]{v_3} \ar@/^1pc/[ll]^{1} }}
&
$\vec{n_i}\deltaeq$
\raisebox{0.75cm}{
\xymatrix{& 
*++[o][F-]{v_1} \ar[dl]^{       i+1} \ar@/^1pc/[dr]^{i+1} & \\
*++[o][F-]{v_2} \ar@/^1pc/[ur]^{i+1} \ar[rr]^{       1} & & 
*++[o][F-]{v_3} \ar[ul]^{       i+1} \ar@/^1pc/[ll]^{1} }}
\\
\\
$\vec{x_{2i}}=$
\raisebox{0.75cm}{
\xymatrix{& 
*++[o][F-]{v_1} \ar@/_0.5pc/[dl]^{2i+1} \ar@/^1.5pc/[dr]^{2i} & \\
*++[o][F-]{v_2} \ar@/^1.5pc/[ur]^{2i+1} \ar[rr]^{       1} & & 
*++[o][F-]{v_3} \ar@/_0.5pc/[ul]^{2i} \ar@/^1pc/[ll]^{1} }}
\quad\quad\quad&
$\vec{x_{2i+1}}=$
\raisebox{0.75cm}{
\xymatrix{& 
*++[o][F-]{v_1} \ar@/_0.5pc/[dl]^{2i+1} \ar@/^1.5pc/[dr]^{2i+2} & \\
*++[o][F-]{v_2} \ar@/^1.5pc/[ur]^{2i+1} \ar[rr]^{       1} & & 
*++[o][F-]{v_3} \ar@/_0.5pc/[ul]^{2i+2} \ar@/^1pc/[ll]^{1} }}
\end{tabular}
\end{center}
\caption{Example of an infinite strictly increasing chain defined by
$\vec{x_0}=\vec{m}^*,\;\vec{x_{i+1}}=(\vec{x_i}\triangledown\vec{n_i})^*$.}
\label{wideningfig}
\end{frameit}
\end{figure}

Originally~\cite{interv},
Cousot and Cousot defined widening over intervals $\overline{\triangledown}$
by:
$$[a,b]\;\overline{\triangledown}\;[c,d]\deltaeq[e,f],$$
where:
$$
\begin{array}{ll}
e\deltaeq\left\{\begin{array}{ll} a &\mbox{if }a\leq c,\\
-\infty\quad&\mbox{elsewhere},\\
\end{array}\right.
\quad\quad&
f\deltaeq\left\{\begin{array}{ll} b &\mbox{if }b\geq d,\\
+\infty\quad&\mbox{elsewhere}\enspace.
\end{array}\right.
\end{array}
$$
The following theorem proves that the sequence computed by
our widening is {\em always} more precise than with
the standard widening over intervals:
\begin{mytheorem}
\label{wideningaccuratethm}
If we have the following iterating sequence:
$$
\begin{array}{ll}
\left\{
\begin{array}{lll}
\vec{x_0}&\deltaeq&\vec{m}^*,\\
\vec{x_{k+1}}&\deltaeq&\vec{x_k}\triangledown(\vec{n_k}^*),
\end{array}
\right.\quad\quad&\left\{
\begin{array}{lll}
\left[y_0,z_0\right]&\deltaeq&\pi_{|v_i}(\vec{m}),\\
\left[y_{k+1},z_{k+1}\right]&\deltaeq&
\left[y_k,z_k\right]\;\overline{\triangledown}\;\pi_{|v_i}(\vec{n_k}),
\end{array}
\right.
\end{array}
$$
then the sequence $(\vec{x_k})_{k\in\mathbb{N}}$ 
is more precise than the sequence $([y_k,z_k])_{k\in\mathbb{N}}$
in the following sense:
$$
\forall k,\;\pi_{|v_i}(\vec{x_k})\subseteq[y_k,z_k]\enspace.
$$
\end{mytheorem}

Remark that the technique, described in 
Cousot and Cousot's PLILP'92 paper \cite{interv2}, for improving
the precision of the standard widening over intervals
$\overline{\triangledown}$
can also be applied to our widening $\triangledown$.
It allows, for instance, deriving a widening that {\it always} gives
better results than a simple sign analysis (which is not the case of
$\triangledown$ nor $\overline{\triangledown}$).
The resulting widening over DBMs will remain more precise than the resulting
widening over intervals.

\subsubsection{Narrowing.}
\mydef{Narrowing operators} were introduced in P. Cousot's thesis 
\cite[\S 4.1.2.0.11]{these}
in order to restore, in a finite time,
some information that may have been lost by widening applications.
We define here a point-wise narrowing operator $\triangle$ by:
$$(\vec{m}\triangle \vec{n})_{ij}\deltaeq
\left\{
\begin{array}{ll}
\vec{n}_{ij} \quad& \mbox{if }\vec{m}_{ij}=+\infty,\\
\vec{m}_{ij} \quad& \mbox{elsewhere}\enspace.
\end{array}
\right.
$$

The following properties prove that $\triangle$ is indeed a narrowing:
\begin{mytheoreml}
\label{narrowingthm}
\item If $\mathcal{D}^0(\vec{n})\subseteq\mathcal{D}^0(\vec{m})$, then
$\mathcal{D}^0(\vec{n})\subseteq\mathcal{D}^0(\vec{m}\triangle\vec{n})\subseteq
\mathcal{D}^0(\vec{m})$.
\item {\it Finite decreasing chain property}:

$\forall \vec{m}$ and for any chain $(\vec{n_i})_{i\in\mathbb{N}}$ 
decreasing for $\trianglelefteqslant$,
the chain defined by:
$$
\left\{
\begin{array}{ll}
\vec{x_0}&\deltaeq\vec{m},\\
\vec{x_{i+1}}&\deltaeq\vec{x_i}\triangle \vec{n_i},
\end{array}
\right.
$$
is decreasing and ultimately stationary.
\end{mytheoreml}

Given a sequence $(\vec{n_k})_{k\in\mathbb{N}}$ such that the chain 
$(\mathcal{D}^0(\vec{n_k}))_{k\in\mathbb{N}}$ 
is decreasing for the $\subseteq$ partial order
(but not $(\vec{n_k})_{k\in\mathbb{N}}$ for the $\trianglelefteqslant$ 
partial order),
one way to ensure the best accuracy
as well as the finiteness of the chain $(\vec{x_k})_{k\in\mathbb{N}}$ 
is to force the closure of the right argument by changing
$\vec{x_{i+1}}=\vec{x_i}\triangle \vec{n_i}$ into
$\vec{x_{i+1}}=\vec{x_i}\triangle (\vec{n_i}^*)$.
Unlike widening, forcing all elements in the chain to be closed
with $\vec{x_{i+1}}=(\vec{x_i}\triangle \vec{n_i})^*$
poses no problem.

\subsubsection{Forget.}
Given a DBM $\vec{m}$ and a variable $v_k$, the \mydef{forget
operator} $\vec{m_{\backslash v_k}}$ computes a DBM where all informations 
about $v_k$ are lost.
It is the opposite of the projection operator $\pi_{|v_k}$.
We define this operator by:
$$
\vec{(m_{\backslash v_k})}_{ij}\deltaeq
\left\{
\begin{array}{ll}
\min(\vec{m}_{ij},\;\vec{m}_{ik}+\vec{m}_{kj})\quad
& \mbox{if }i\neq k\mbox{ and }j\neq k,\\
0&\mbox{if }i=j=k,\\
+\infty & \mbox{elsewhere}\enspace.
\end{array}
\right.
$$

The $\mathcal{V}^0$-domain of $\vec{m_{\backslash v_k}}$ is obtained by 
projecting $\mathcal{D}^0(\vec{m})$ on the subspace orthogonal to
$\mathbb{I}\overrightarrow{\vec{v_k}}$, 
and then extruding the result in the direction of 
$\overrightarrow{\vec{v_k}}$:
\begin{mytheorem}
\label{forgetthm}
$\begin{array}{l}
\mathcal{D}^0(\vec{m_{\backslash v_k}})\;=\\
\quad\{(x_2,\ldots,x_n)\in\mathbb{I}^{n-1}\;|\;\exists x\in\mathbb{I},
(x_2,\ldots,x_{k-1},x,x_{k+1},\ldots,x_n)\in\mathcal{D}^0(\vec{m})\}.
\end{array}$

\end{mytheorem}

\subsubsection{Guard.}
Given an arithmetic equality or inequality $g$ over 
$\{v_2,\ldots,v_n\}$---which we call a \mydef{guard}---and 
a DBM $\vec{m}$, the
\mydef{guard transfer function} tries to find a new DBM $\vec{m_{(g)}}$ the 
$\mathcal{V}^0$-domain of which is
$\{s\in\mathcal{D}^0(\vec{m})\;|\;s\mbox{ satisfies }g\}$.
Since this is, in general, impossible, we will only try to have:
\begin{mytheorem}
\label{guardthm}
$\mathcal{D}^0(\vec{m_{(g)}})\supseteq
\{s\in\mathcal{D}^0(\vec{m})\;|\;s\mbox{ satisfies }g\}$.
\end{mytheorem}

Here is an example definition:
\begin{mydefinitionl}
\label{guarddef}
\item If $g=(v_{j_0}-v_{i_0}\leq c)$ with $i_0\neq j_0$,
then:
$$
\vec{(m_{(v_{j_0}-v_{i_0}\leq c)})}_{ij}\deltaeq
\left\{
\begin{array}{ll}
\min(\vec{m}_{ij},c)\quad&\mbox{if }i=i_0\mbox{ and }j=j_0,\\
\vec{m}_{ij}&\mbox{elsewhere}\enspace.
\end{array}
\right.
$$
The cases $g=(v_{j_0}\leq c)$ and $g=(-v_{i_0}\leq c)$ are
settled by choosing respectively $i_0=1$ and $j_0=1$.
\medskip

\item If $g=(v_{j_0}-v_{i_0}=c)$ with $i_0\neq j_0$,
then:
$$
\vec{m_{(v_{j_0}-v_{i_0}=c)}}\deltaeq
\vec{(m_{(v_{j_0}-v_{i_0}\leq c)})_{(v_{i_0}-v_{j_0}\leq -c)}}\enspace.
$$
The case $g=(v_{j_0}=c)$ is a special case where $i_0=1$.
\medskip

\item In all other cases, we simply choose:
$$
\vec{m_{(g)}}\deltaeq\vec{m}\enspace.
$$
\end{mydefinitionl}

In all but the last---general---cases, the guard transfer function is exact.

\subsubsection{Assignment.}

An \mydef{assignment} $v_k\leftarrow e(v_2,\ldots,v_n)$ 
is defined by a variable $v_k$ and an arithmetic 
expression $e$ over $\{v_2,\ldots,v_n\}$.

Given a DBM $\vec{m}$ representing all possible values that can take the
variables set $\{v_2,\ldots,v_n\}$ at a program point, 
we look for a DBM, denoted by $\vec{m_{(v_k\leftarrow e)}}$,
representing the possibles values of the same variables set after
the assignment $v_k\leftarrow e$.
This is not possible in the general case, so the 
\mydef{assignment transfer function} 
will only try to find an upper approximation of this set:

\begin{mytheorem}
\label{assignmentthm}
$\begin{array}{l}\mathcal{D}^0(\vec{m_{(v_k\leftarrow e)}})\;\supseteq\\
\quad\{(x_2,\ldots,x_{k-1},e(x_2,\ldots,x_n),x_{k+1},\ldots,
x_n)\;|\;(x_2,\ldots,x_n)\in\mathcal{D}^0(\vec{m})\}\enspace.\end{array}$
\end{mytheorem}

For instance, we can use the following definition
for $\vec{m_{(v_{i_0}\leftarrow e)}}$:

\begin{mydefinitionl}
\label{assignmentdef}
\item If $e=v_{i_0}+c$, then:
$$
\vec{(m_{(v_{i_0}\leftarrow v_{i_0}+c)})}_{ij}\deltaeq
\left\{
\begin{array}{ll}
\vec{m}_{ij}-c\quad&\mbox{if }i=i_0,j\neq j_0,\\
\vec{m}_{ij}+c&\mbox{if }i\neq i_0,j=j_0,\\
\vec{m}_{ij}&\mbox{elsewhere}\enspace.
\end{array}
\right.
$$
\medskip

\item If $e=v_{j_0}+c$ with $i_0\neq j_0$,
then we use the forget operator and the guard transfer function:
$$
\vec{m_{(v_{i_0}\leftarrow v_{j_0}+c)}}\deltaeq
\vec{((m_{\backslash v_{i_0}})_{
(v_{i_0}-v_{j_0}\leq c)})_{(v_{j_0}-v_{i_0}\leq -c)}}\enspace.
$$
The case $e=c$ is a special case where we choose $j_0=1$.
\medskip

\item In all other cases, we use a standard interval arithmetic 
to find an interval $[-e^-,e^+]$, $e^+,e^-\in\overline{\mathbb{I}}$
such that 
$$[-e^-,e^+]\;\supseteq\;e(\pi_{v_2}(\vec{m}),\ldots,\pi_{v_n}(\vec{m}))$$
and then we define:
$$
\vec{(m_{(v_{i_0}\leftarrow e)})_{ij}}\deltaeq
\left\{
\begin{array}{ll}
e^+&\mbox{if $i=1$ and $j=i_0$},\\
e^-&\mbox{if $j=1$ and $i=i_0$},\\
\vec{(m_{\backslash v_{i_0}})_{ij}}\quad&\mbox{elsewhere}\enspace.
\end{array}
\right.
$$
\end{mydefinitionl}

In all but the last---general---cases, the assignment transfer function is 
exact.

\subsubsection{Comparison with the Abstract Domain of Intervals.}
Most of the time, the precision of numerical abstract domains can only
be compared experimentally on example programs (see Section 6 for such 
an example).
However, we claim that the DBM domain {\em always} performs better than
the domain of intervals. 

To legitimate this assertion, we compare informally the effect of all abstract
operations in the DBM and in the interval domains.
Thanks to Theorems \ref{intersectionthm} and \ref{lubthm}.2, and
Definitions \ref{guarddef} and \ref{assignmentdef}, the intersection
and union abstract operators and the guard and assignment transfer
functions are more precise than their interval counterpart.
Thanks to Theorem \ref{wideningaccuratethm}, approximate fixpoint computation
with our widening $\triangledown$ is always more accurate than with
the standard widening over intervals $\overline{\triangledown}$ and one
could prove easily that each iteration with our narrowing is
more precise than with the standard narrowing over intervals.
This means that {\em any} abstract semantics based on the operators and 
transfer functions we defined is {\em always} more precise than the
corresponding interval-based abstract semantics.

\section{Lattice Structures}

In this section, we design two lattice structures: one on the set of DBMs
and one on the set of closed DBMs.
The first one is useful to analyze fixpoint transfer between
abstract and concrete semantics and the second one
allows us to design a meaning function---or 
even a Galois Connection---linking the set of
abstract $\mathcal{V}^0$-domains to the concrete lattice
$\mathcal{P}(\{v_2,\ldots,v_n\}\mapsto\mathbb{I})$, following the
abstract interpretation framework
described in Cousot and Cousot's POPL'79 paper \cite{semdesign}.

\subsubsection{DBM Lattice.}
The set $\mathcal{M}$ of DBMs, 
together with the \mydef{order relation} $\trianglelefteqslant$
and the point-wise \mydef{least upper bound} $\vee$ and 
\mydef{greatest lower bound}
$\wedge$, is almost a lattice. 
It only needs a \mydef{least element $\bot$}, so we
extend $\trianglelefteqslant$, $\vee$ and $\wedge$ to
$\mathcal{M}_{\bot}=\mathcal{M}\cup\{\bot\}$ in an obvious way to get
$\sqsubseteq$, $\sqcup$ and $\sqcap$.
The \mydef{greatest element $\top$} is the DBM with all its coefficients 
equal to $+\infty$.

\begin{mytheoreml}
\item $(\mathcal{M}_{\bot},\sqsubseteq,\sqcap,\sqcup,\bot,\top)$ is a lattice.
\item This lattice is complete if $(\mathbb{I},\leq)$ is complete
($\mathbb{Z}$ or $\mathbb{R}$, but not $\mathbb{Q}$).
\end{mytheoreml}

There are, however, two problems with this lattice.
First, we cannot easily assimilate this lattice to a sub-lattice of
$\mathcal{P}(\{v_2,\ldots,v_n\}\mapsto\mathbb{I})$ as two different
DBMs can have the same $\mathcal{V}^0$-domain.
Then, the least upper bound operator $\sqcup$ is not the most
precise upper approximation of the union of 
two $\mathcal{V}^0$-domains because we do 
not force the arguments to be closed.

\subsubsection{Closed DBM Lattice.}
To overcome these difficulties, we build another lattice based
on closed DBMs.
First, consider the set $\mathcal{M}^*_{\bot}$ of closed DBMs
$\mathcal{M}^*$ with a \mydef{least element $\bot^*$} added.
Now, we define a \mydef{greatest element $\top^*$}, 
a \mydef{partial order relation $\sqsubseteq^*$},
a \mydef{least upper bound $\sqcup^*$} and 
a \mydef{greatest lower bound $\sqcap^*$} in $\mathcal{M}^*_{\bot}$ by:
\vspace*{0.2cm}

$
\vec{\top^*}_{ij}\deltaeq
\left\{
\begin{array}{ll}
0&\mbox{if }i=j,\\
+\infty\quad&\mbox{elsewhere}\enspace.
\end{array}\right.
$
\vspace*{0.2cm}

$
\vec{m}\sqsubseteq^*\vec{n}\deltaiff
\left\{
\begin{array}{ll}
\mbox{either }&\vec{m}=\bot^*,\\
\mbox{or }&\vec{m}\neq\bot^*, \vec{n}\neq \bot^*\mbox{ and }
\vec{m}\trianglelefteqslant\vec{n}\enspace.
\end{array}\right.
$
\vspace*{0.2cm}

$
\vec{m}\sqcup^*\vec{n}\deltaeq
\left\{
\begin{array}{ll}
\vec{m}&\mbox{if }\vec{n}=\bot^*,\\
\vec{n}&\mbox{if }\vec{m}=\bot^*,\\
\vec{m}\vee\vec{n}\quad&\mbox{elsewhere}\enspace.
\end{array}\right.
$
\vspace*{0.2cm}

$
\vec{m}\sqcap^*\vec{n}\deltaeq
\left\{
\begin{array}{ll}
\vec{\bot^*}&\mbox{if }\vec{m}=\bot^*\mbox{ or }\vec{n}=\bot^*
\mbox{ or }\mathcal{D}^0(\vec{m}\wedge\vec{n})=\emptyset,\\
(\vec{m}\wedge\vec{n})^*\quad&\mbox{elsewhere}\enspace.
\end{array}\right.
$

Thanks to Theorem \ref{closurethm}.1, every non-empty $\mathcal{V}^0$-domain
has a unique 
representation in $\mathcal{M}^*$; $\bot^*$ is the representation for
the empty set.
We build a \mydef{meaning function $\gamma$} which is an extension
of $\mathcal{D}^0(\cdot)$ to $\mathcal{M}^*_{\bot}$:
$$
\gamma(\vec{m})\deltaeq
\left\{
\begin{array}{ll}
\emptyset&\mbox{if }\vec{m}=\bot^*,\\
\mathcal{D}^0(\vec{m})\quad&\mbox{elsewhere}\enspace.
\end{array}\right.
$$

\begin{mytheoreml}
\item $(\mathcal{M}^*_{\bot},\sqsubseteq^*,\sqcap^*,\sqcup^*,\bot^*,\top^*)$ 
is a lattice and $\gamma$ is one-to-one.
\item If $(\mathbb{I},\leq)$ is complete, this lattice is complete and
$\gamma$ is meet-preserving:\\ $\gamma(\bigsqcap^*X)=\bigcap\{\gamma(x)\;|\;
x\in X\}$. 
We can---according to Cousot and Cousot \cite[Prop. 7]{fixpoint}---build 
a canonical \mydef{Galois Insertion}:
$$\mathcal{P}(\{v_2,\ldots,v_n\}\mapsto\mathbb{I})\;
\galoiS{\alpha}{\gamma}\;
\mathcal{M}^*_{\bot}$$
where the \mydef{abstraction function $\alpha$} is defined by:\\
$\alpha(D)=\bigsqcap^*\;\{\;m\in\mathcal{M}^*_{\bot}
\;|\;D\subseteq\gamma(m)\;\}$.

\end{mytheoreml}

The $\mathcal{M}^*_{\bot}$ lattice features a nice meaning function
and a precise union approximation;
thus, it is tempting to force all our operators and transfer functions to 
live in $\mathcal{M}^*_{\bot}$ by forcing closure on their result.
However, we saw this does not work for widening, so fixpoint computation
{\em must} be performed in the $\mathcal{M}_{\bot}$ lattice.

\section{Results}

The algorithms on DBMs presented here have been implemented in \textsf{OCaml}
and used to perform forward analysis on toy---yet 
Turing-equivalent---imperative 
and parallel languages with only numerical variables and no procedure.

We present here neither the concrete and abstract semantics,
nor the actual forward analysis algorithm used for our analyzers.
They follow exactly the abstract interpretation scheme 
described in Cousot and Cousot's
POPL'79 paper \cite{semdesign} and Bourdoncle's FMPA'93 paper \cite{chaotic} 
and are detailed in the author's MS thesis \cite{dea}.
Theorems \ref{emptythm}, \ref{testthm}, \ref{intersectionthm}, \ref{lubthm},
\ref{guardthm} and \ref{assignmentthm} prove
that all the operators and transfer functions we defined are indeed
abstractions on the domain of DBMs
of the usual operators and transfer functions on the concrete
domain $\mathcal{P}(\{v_2,\ldots,v_n\}\mapsto\mathbb{I})$, which, 
as shown by Cousot and Cousot~\cite{semdesign}, is sufficient
to prove soundness for analyses.

\subsubsection{Imperative Programs.}
Our toy forward analyzer for imperative language follows almost
exactly the analyzer described in 
Cousot and Halbwachs's POPL'78 paper \cite{poly}, except that the
abstract domain of polyhedra has been replaced by our DBM-based domain.
We tested our analyzer on the well-known Bubble Sort and Heap Sort
algorithms and managed to prove automatically that they
do not produce out-of-bound error while accessing array elements.
Although we did not find as many invariants as Cousot and Halbwachs
for these two examples, it was sufficient to prove the correctness.
We do not detail these common examples here for the sake of brevity.

\subsubsection{Parallel Programs.}
Our toy analyzer for parallel language allows analyzing a fixed
set of processes running concurrently and communicating through global
variables. We use the well-known {\em nondeterministic interleaving} 
method in order to analyze all possible control flows.
In this context, we managed to prove automatically
that the Bakery algorithm, introduced in 1974 by Lamport 
\cite{bakery}, for synchronizing two parallel 
processes never lets the two processes
be at the same time in their critical sections.
We now detail this example.

\subsubsection{The Bakery Algorithm.}
After the initialization of two global shared variables $y1$ and $y2$, 
two processes $p1$ and $p2$ are spawned. They synchronize through the 
variables $y1$ and $y2$, representing the priority of $p1$ and $p2$, so 
that only one process at a time can enter its {\it critical section}
(Figure \ref{bakeryalg}).

\begin{figure}
\begin{frameit}
\begin{center}
\textsf{
\begin{tabular}{l}
$y1=0$; $y2=0$;\\
\\
{\bf (p1)}\\
\hline
\quad {\bf while true do}\\
\quad\quad $y1=y2+1$;\\
\quad\quad {\bf while} $y2\neq 0$ {\bf and} $y1>y2$ {\bf do done};\\
\quad\quad {\it - - - critical section - - -}\\
\quad\quad $y1=0$;\\
\quad {\bf done}\\
\\
{\bf (p2)}\\
\hline
\quad {\bf while true do}\\
\quad\quad $y2=y1+1$;\\
\quad\quad {\bf while} $y1\neq 0$ {\bf and} $y2\geq y1$ {\bf do done};\\
\quad\quad {\it - - - critical section - - -}\\
\quad\quad $y2=0$;\\
\quad {\bf done}\\
\end{tabular}}
\end{center}
\caption{Pseudo-code for the Bakery algorithm.}
\label{bakeryalg}
\end{frameit}
\end{figure}

\begin{figure}
\begin{frameit}
\begin{center}
\begin{tabular}{cc}
\footnotesize
\begin{picture}(55,60)
\node[Nmarks=i,iangle=145](a0)(20,50){$0$}
\node(a1)(20,30){$1$}
\node(a2)(20,10){$2$}
\drawedge(a0,a1){$y1\leftarrow y2+1$}
\drawedge(a1,a2){$y2=0$ or $y1\leq y2$}
\drawloop[loopangle=0,loopdiam=5](a1){$y2\neq 0$ and $y1>y2$}
\drawloop[loopangle=0,loopdiam=5,dash={0.2 0.5}0](a2){{\it critical section}}
\drawbcedge[ELpos=40](a2,10,0,a0,10,60){$y1\leftarrow 0$}
\end{picture}
\quad\quad&
\footnotesize
\begin{picture}(55,60)
\node[Nmarks=i,iangle=145](a0)(20,50){$a$}
\node(a1)(20,30){$b$}
\node(a2)(20,10){$c$}
\drawedge(a0,a1){$y2\leftarrow y1+1$}
\drawedge(a1,a2){$y1=0$ or $y2<y1$}
\drawloop[loopangle=0,loopdiam=5](a1){$y1\neq 0$ and $y2\geq y1$}
\drawloop[loopangle=0,loopdiam=5,dash={0.2 0.5}0](a2){{\it critical section}}
\drawbcedge[ELpos=40](a2,10,0,a0,10,60){$y2\leftarrow 0$}
\end{picture}
\\
$(p1)$&$(p2)$
\end{tabular}
\end{center}
\caption{Control flow graphs of processes $p1$ and $p2$
in the Bakery algorithm.}
\label{bakerygraph}
\end{frameit}
\end{figure}
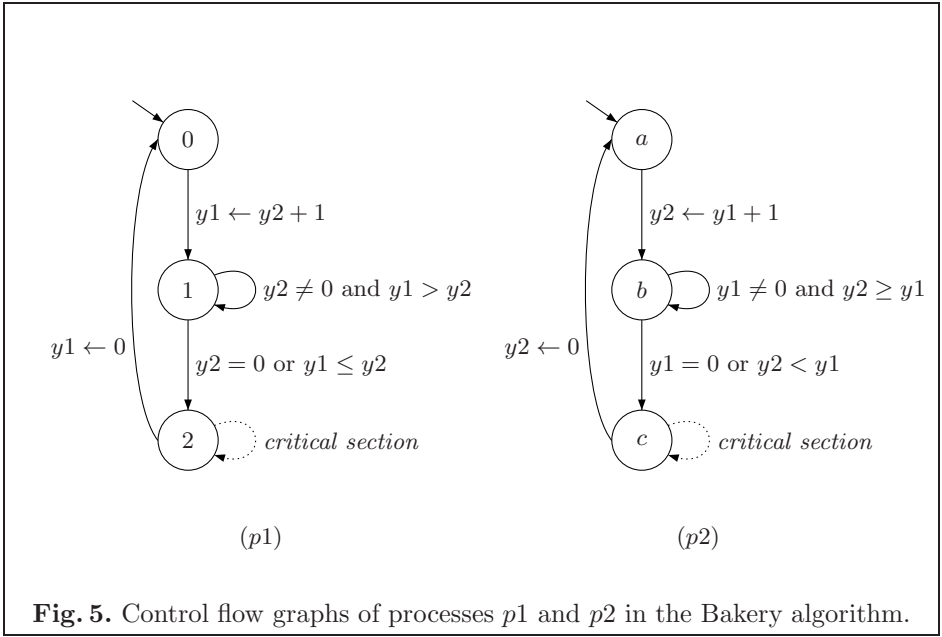

Our analyzer for parallel processes is fed with the initialization
code ($y1=0$; $y2=0$) and the control flow graphs for
$p1$ and $p2$ (Figure \ref{bakerygraph}).
Each control graph is a set of control point nodes and
some edges labeled with 
either an \mydef{action} performed when the edge is taken
(the assignment $y1\leftarrow y2+1$, for example)
or a \mydef{guard} imposing a condition for taking the edge
(the test $y1\neq 0$, for example).

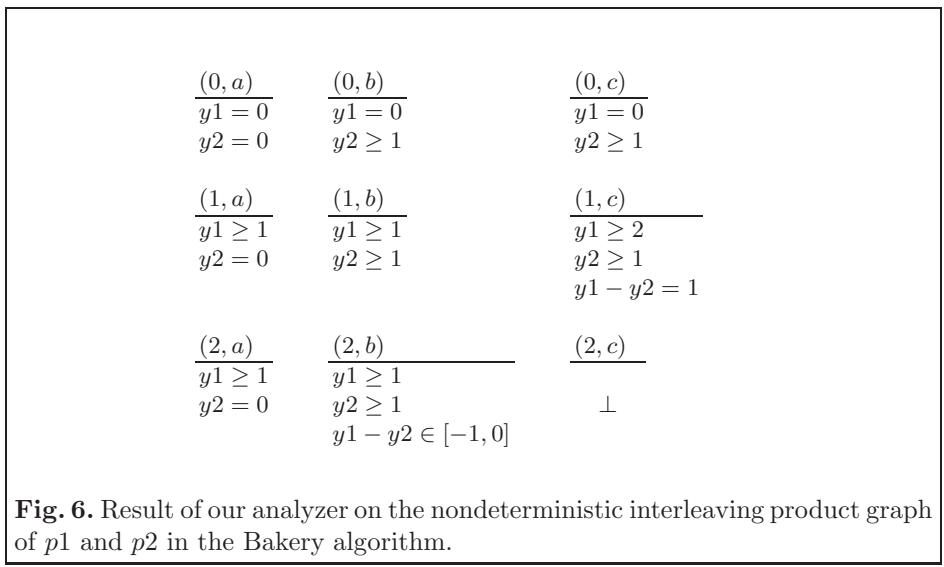
\begin{figure}
\begin{frameit}
\begin{center}
$
\begin{array}{lll}
\begin{array}{l}
(0,a)\\
\hline
y1=0\\
y2=0\\
\end{array}\quad\quad
&
\begin{array}{l}
(0,b)\\
\hline
y1=0\\
y2\geq 1\\
\end{array}\quad\quad
&
\begin{array}{l}
(0,c)\\
\hline
y1=0\\
y2\geq 1\\
\end{array}\quad\quad
\\\\
\begin{array}{l}
(1,a)\\
\hline
y1\geq 1\\
y2=0\\
\\
\end{array}\quad\quad
&
\begin{array}{l}
(1,b)\\
\hline
y1\geq 1\\
y2\geq 1\\
\\
\end{array}\quad\quad
&
\begin{array}{l}
(1,c)\\
\hline
y1\geq 2\\
y2\geq 1\\
y1-y2=1\\
\end{array}\quad\quad
\\\\
\begin{array}{l}
(2,a)\\
\hline
y1\geq 1\\
y2=0\\
\\
\end{array}\quad\quad
&
\begin{array}{l}
(2,b)\\
\hline
y1\geq 1\\
y2\geq 1\\
y1-y2 \in [-1,0]\\
\end{array}\quad\quad
&
\begin{array}{l}
(2,c)\\
\hline
\\
\quad\bot\quad\\
\\
\end{array}
\end{array}
$
\end{center}
\caption{Result of our analyzer on the nondeterministic interleaving product
graph of $p1$ and $p2$ in the Bakery algorithm.}
\label{bakeryresult}
\end{frameit}
\end{figure}

The analyzer then computes the nondeterministic interleaving of $p1$ and $p2$
which is the product control flow graph.
Then, it computes iteratively the abstract invariants holding at each
product control point.
It outputs the invariants shown in Figure \ref{bakeryresult}.

The state $(2,c)$ is never reached, which means that $p1$ and $p2$ cannot
be at the same time in their critical section. This proves the
correctness of the Bakery algorithm.
Remark that our analyzer also discovered some non-obvious
invariants, such as $y1=y2+1$ holding in the $(1,c)$ state.

\section{Extensions and Future Work}

\subsubsection{Precision improvement.}
In our analysis, we only find a coarse set of the invariants held in a 
program since finding {\em all} invariants of the form
$(x-y\leq c)$ and $(\pm x\leq c)$ for all programs is 
non-computable.
Possible losses of precision have three causes: non-exact union,
widening in loops and non-exact assignment and guard transfer functions.

We made crude approximations in the last---general---case of 
Definitions \ref{guarddef} and \ref{assignmentdef} and there is room for
improving assignment and guard transfer functions,
even though exactness is impossible.
When the DBM lattices are complete, there exists most precise transfer 
functions such that Theorems \ref{guardthm} and \ref{assignmentthm} hold, 
however these functions may be difficult to compute.

\subsubsection{Finite Union of $\mathcal{V}^0$-domains.}
One can imagine to represent finite unions of $\mathcal{V}^0$-domains,
using a finite set of DBMs instead of a single one as abstract state.
This allows an exact union operator but
it may lead to memory and time cost explosion as abstract
states contain more and more DBMs, so one may need from time to time to
replace a set of DBMs by their union approximation.

The model-checker community has also developed specific structures
to represent finite unions of $\mathcal{V}$-domains, that are less costly than
sets. {\em Clock-Difference Diagrams} 
(introduced in 1999 by Larsen, Weise, Yi and Pearson \cite{CDD}) 
and {\em Difference Decision Diagrams} 
(introduced in M{\o}ller, Lichtenberg, Andersen and Hulgaard's
CSL'99 paper \cite{DDD})
are tree-based structures made compact
thanks to the sharing of isomorphic sub-trees; however existence of
normal forms for such structures is only a conjecture at the time
of writing and only local or path reduction algorithms exist.
One can imagine adapting such structures to abstract interpretation the
way we adapted DBM in this paper.

\subsubsection{Space and Time Cost Improvement.}
Space is often a big concern in abstract interpretation.
The DBM representation we proposed in this paper has a
fixed $\mathcal{O}(n^2)$ memory cost---where $n$ is the number
of variables in the program.
In the actual implementation, we decided to use the graph
representation---or hollow matrix---which stores only
edges with a finite weight and observed a great space gain
as most DBMs we use have many $+\infty$.
Most algorithms are also faster on
hollow matrices and we chose to use the more complex, but more
efficient, {\em Johnson} shortest-path closure algorithm---described 
in Cormen, Leiserson and Rivest's textbook \cite[\S 26.3]{CLR}---instead of 
the {\em Floyd-Warshall} algorithm.

Larsen, Larsson, Pettersson and Yi's RTSS'97 paper \cite{DBM} presents a
\mydef{minimal form algorithm} which
finds a DBM with the fewest finite edges representing
a given $\mathcal{V}^0$-domain.
This minimal form could be useful for memory-efficient
storing, but cannot be used for direct computation with algorithms
requiring closed DBMs.

\subsubsection{Representation Improvement.}
The invariants we manipulate are, in term of precision and complexity,
between interval and polyhedron analysis.
It is interesting to look for domains allowing the representation of
more forms of invariants than DBMs in order to increase the granularity 
of numerical domains.
We are currently working on an improvement of DBMs that allows us
to represent, with a small time and space complexity overhead,
invariants of the form $(\pm x \pm y \leq c)$.

\section{Conclusion}

We presented in this paper a new numerical abstract domain inspired
from the well-known domain of intervals and the Difference-Bound Matrices.
This domain allows us to manipulate invariants of the form 
$(x-y\leq c)$, $(x\leq c)$ and $(x\geq c)$ with a
$\mathcal{O}(n^2)$ worst case memory cost per
abstract state and $\mathcal{O}(n^3)$ worst case time cost per abstract 
operation (where $n$ is the number of variables in the program).

Our approach made it possible for us
to prove the correctness of some non-trivial algorithms 
beyond the scope of interval analysis, for a much smaller cost than polyhedron
analysis.
We also proved that this analysis always gives better results than interval
analysis, for a slightly greater cost.

\subsubsection{Acknowledgments.}
I am grateful to J. Feret, C. Hymans, D. Monniaux, P. Cousot, O. Danvy
and the anonymous referees for their helpful comments and suggestions.


\bibliographystyle{plain}
\bibliography{bibarticle}

\end{document}